\documentclass[pss,fleqn,embeddedheads]{w-art}
\usepackage{times}
\usepackage{w-thm}
\usepackage[]{graphicx}
\begin{document}
\DOIsuffix{theDOIsuffix}
\Volume{XX}
\Issue{1}
\Copyrightissue{01}
\Month{01}
\Year{2004}
\pagespan{1}{}
\Receiveddate{\sf zzz} \Reviseddate{\sf zzz} \Accepteddate{\sf
zzz} \Dateposted{\sf zzz}
\subjclass[pacs]{75.70.Cn,75.50.Pp,75.30.Et}



\title[Band-gap and chemical effects on the electronic structure and exchange constants in magnetic semiconductors digital alloys]{Electronic structure and exchange constants in magnetic semiconductors digital alloys: chemical and band-gap effects}


\author[S. Picozzi]{S. Picozzi\footnote{Corresponding
     author: e-mail: {\sf silvia.picozzi@aquila.infn.it}, Phone: +39\,0862\,43\,3055, Fax:
     +39\,0862\,43\,3033}\inst{1}}
\address[\inst{1}]{CNR-INFM SENSOR-CASTI Regional Lab., Coppito, L'Aquila, Italy}
\author[M. Le\v{z}ai\'{c}]{M. Le\v{z}ai\'{c}\inst{2}}
\address[\inst{2}]{Institut f\"ur Festk\"orperforschung, Forschungszentrum Juelich, Germany}

\author[]{S. Bl\"ugel\inst{2}}
\begin{abstract}
First-principles simulations have been performed for [001]-ordered Mn/Ge and Mn/GaAs ``digital alloys", focusing on the effects of {\em i}) a larger band-gap and {\em ii}) a different semiconducting host on the electronic structure of the magnetic semiconductors of interest. Our results for the exchange constants in Mn/Ge, evaluated using a frozen-magnon scheme, show that a larger band-gap tends to give a stronger nearest-neighbor ferromagnetic coupling and an overall enhanced in-plane ferromagnetic coupling even for longer-ranged coupling constants. As for the chemical effects on the exchange constants, we show that  Mn/GaAs shows a smaller nearest-neighbor ferromagnetic coupling than Mn/Ge, but exchange constants for higher Mn-Mn distance show an overall increased ferromagnetic behavior in Mn/GaAs. As a result, from the magnetic-coupling point of view, the two systems  behave on average rather similarly.
\end{abstract}
\maketitle                   




\renewcommand{\leftmark}
{S. Picozzi et al.: Electronic structure and  exchange constants in  magnetic semiconductors digital alloys}

\section{Introduction}

Diluted magnetic semiconductors (DMS), {\em i.e.} mainstream semiconductors (SC) doped with small amounts of transition metal impurities, represent a class of materials with potentially relevant applications in modern spintronics. The discovery of ferromagnetism (FM) persisting to temperatures up to 
$\leq$170 K in the DMS prototype\cite{ohno}, {\em i.e.} GaMnAs, motivated a wealth of studies focused on Mn-doped semiconductors\cite{review}. However,  crucial issues are still fully open to debate, such as {\em i}) the origin of ferromagnetism in DMS, from the theoretical point of view,  {\em ii}) a careful control over the material quality during growth, from the experimental point of view and  {\em iii}) the choice of a material that might show a Curie temperature higher than room temperature, from the technological point of view. 
 One of the key problems in DMS is the low solubility of Mn in semiconductors and severely out-of-equilibrium techniques (such as Molecular Beam Epitaxy, MBE)
 have to be used to successfully achieve a DMS growth  without precipitates or large concentration of defects. However, many groups succeeded in doping semiconductors with large Mn concentrations by means of ``digital alloys" (DA),  artificial heterostructures that typically consists of
a Mn plane with a nominal 50$\%$ concentration alternated with 10-15 layers of SC\cite{furdynadig,altridig}. 

In addition, group-IV-based spintronics is extremely appealing, due to high compatibility with mainstream
Si-based technology and this has motivated the recent interests on Mn-doped Ge 
DMS\cite{steve,stroppa,zhao,kett}. In particular, 
several studies ascertained the ferromagnetism of MBE-grown homogeneously-doped MnGe alloys up to temperatures of $\sim$ 110 K.\cite{weitering} In addition, an extensive characterization was performed on MnGe DMS grown by means of ion-implantation, which revealed the coexistence of Mn$_5$Ge$_3$ nanoclusters dispersed in the SC matrix\cite{lot}. 

From the theoretical point of view, some of us investigated Mn/Ge digital alloys\cite{digiale}, focusing on the electronic structure and on the coupling between two Mn planes as a function of the SC spacer thickness. In particular, the results showed that an efficient 
interplanar magnetic coupling vanishes for spacers thicker than 5-7 layers and this spatial extension perfectly matches the hole-distribution, consistent with the commonly accepted idea that FM in DMS occurs through a hole-mediated mechanism.\cite{review}

Despite large interests in DA, the calculation of their exchange constants has - to our knowledge - not been carefully tackled yet. Therefore, we here consider both  Mn/GaAs and Mn/Ge DA, focusing on the evaluation from first-principles  of their coupling constants. In particular,  we will examine two relevant effects on these calculated constants: {\em i}) {\em chemical} effects, as given by the local coordination of Mn with As vs Ge atoms 
in Mn/GaAs and Mn/Ge DA, respectively, and {\em ii}) {band-gap} effects, as obtained by artificially changing the amplitude of the forbidden energy-gap in the SC host, by means of the so-called  ``LDA+U" approach.
The paper is organized as follows: after reporting some technicalities in Sec.\ref{comp} and after recalling the main hybridization mechanism that occurs upon Mn doping of the hosting SC (Sec.\ref{azlev}), we will move to the discussion of the effects on the electronic structure and on the exchange constants induced by a larger band-gap (Sec.\ref{band-gap}) and by a different SC host (Sec.\ref{chemical}).
 
\section{Computational and structural details}
\label{comp}

\begin{SCfigure}[4][b]
\includegraphics[width=.43\textwidth]{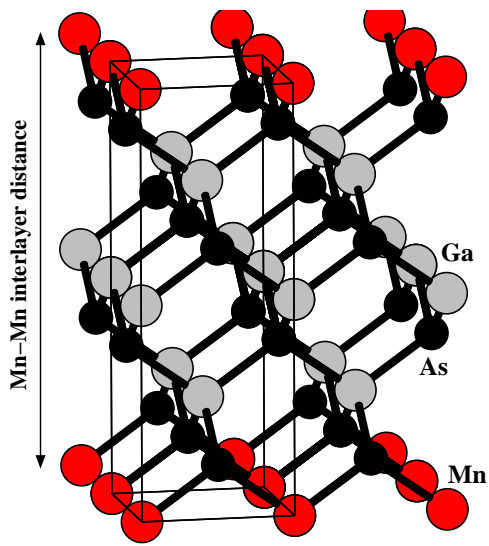}
\caption{Schematic plot of [001]-ordered Mn/GaAs DA, along with the chosen unit cell. Mn, Ga and As atoms are denoted with red, grey and black spheres, respectively. A similar cell has been used for Mn/Ge.} 
\label{fig:cell}
\end{SCfigure}

We report in Fig.\ref{fig:cell} a schematic plot to visualize the tetragonal unit cell, with Bravais vectors {\bf a}= (a/$\sqrt{2}$,0,0), {\bf b}= (0,a/$\sqrt{2}$,0) and {\bf c}= (0,0,2a), where a is the SC lattice constant.
Therefore, the systems consist of a [001]-ordered ``superlattice" with 7 monolayers of semiconductors and one intercalated Mn plane. The nominal concentration of Mn in the SC host is therefore equal to 12.5~$\%$ and 25~$\%$ in MnGe and GaMnAs, respectively. As pointed out before, despite solubility problems that prevent a large doping of Mn in the SC host homogeneously,  the growth of a single Mn plane with very high concentrations can be much better controlled and, therefore, the Mn concentration  in the unit cell  is physically meaningful.   From the experimental point of view, the thickness of the semiconducting spacer is generally larger ($\sim$ 10-15 layers)\cite{furdynadig}; however, it was shown in Ref.\cite{digiale} that a thickness of 7 SC layers 
between two Mn monolayers in nearby unit-cells is sufficiently large to completely decouple their exchange interaction, so that {\em interplanar} exchange interactions are almost negligible and only 
{\em intraplanar} exchange interactions are relevant.
The  lattice constants are chosen according to their experimental values \cite{LB}: 5.653 \AA$\:$ and 5.658 \AA$\:$ for GaAs
and Ge, respectively. Since their difference is very small ({\em i.e.} less than  0.1 $\%$), this also allows to compare  the exchange constants at ``fixed" volume - or ``fixed"  bond-lengths -
and to ascribe the changes between the two systems to chemical effects, 
since structural effects (that might of course also affect the results) act in a similar way in both compounds.
Previous first-principles simulations using the same FLAPW methodology showed that atomic  relaxations of MnGe\cite{stroppa}
and GaMnAs\cite{ajfgamnas} are rather small ({\em i.e.} less than 1.8 $\%$ on the Mn-Ge bond length) and do not sensibly affect the resulting electronic structure. Therefore,  in the present work the atomic positions
were kept fixed to 
those of the ideal zincblende lattice.

As for the computational details, first-principles
calculations have been performed within the spin-density functional theory (DFT) using the generalized gradient approximation (GGA) to the exchange--correlation potential \cite{pbe} and the full-potential linearized augmented plane wave (FLAPW) \cite{flapw} method in the FLEUR implementation 
\cite{fleur}.
 
 Muffin-tin radii were chosen as 2.25 (2.31) a.u. for Mn, Ge and As (Ga), whereas the wave-function expansion was carried out
 with a cut-off of 3.55 (a.u.)$^{-1}$. During the self-consistency cicle, the Brillouin zone (BZ) was sampled
 using 42 special {\bf k}-points, chosen according to the Monkhorst-Pack scheme\cite{mp} in the irreducible BZ.
 
 As for the calculation of the exchange constants , we used the force theorem to approximate the total-energy differences 
of frozen magnons by the differences of the sums of eigenvalues and 
we then deduced the magnon dispersion relations on a set of spin-spiral vectors from the irreducible wedge of the BZ. A subsequent Fourier transformation 
yielded the real-space exchange constants J$_{ij}$ of the Heisenberg model.\cite{sandratskii,mrj}
The calculation was performed on a dense mash of 1536 spin-spirals with different wave-vectors {\bf q} in the full Brillouin zone
and a set of 576 k-points in the full BZ was used.
 This choice of numerical parameters was performed after extensive tests on the sets of both {\bf k}- and {\bf q}-points: the errors
 on the $J_{ij}$ exchange constants are estimated to be less than 1$\%$ with respect to larger sets
 of {\bf k}-points and spin-spiral wave-vectors.
 
 In order to investigate the effects on the exchange constants of having  a larger band-gap (both in the Ge host and, as a consequence, in the minority spins of Mn/Ge - see below) , 
we have used an artifact based on the LDA+U (or GGA+U) scheme\cite{ldaU} as implemented in FLEUR\cite{shick}. According to this approach, an orbital-dependent potential following a Hubbard--like model is combined with the traditional LDA scheme. 
This method was traditionally developed  to treat the electronic structure of highly--correlated materials,
but it has also been successfully applied to semiconductors \cite{antonov:gb} to better take into account quasi--particle effects that are neglected in a bare-LDA approach. Here, we applied a U on the Ge $p$ states, in order to open a gap (forcing the density matrix to be fixed so as to fully occupy the $p$ shell); in particular, a value of U = 2 eV gave a Ge band-gap in excellent agreement
with experiments ($E_g \sim$  0.65 eV). 

\begin{SCfigure}
\includegraphics[width=.4\textwidth]{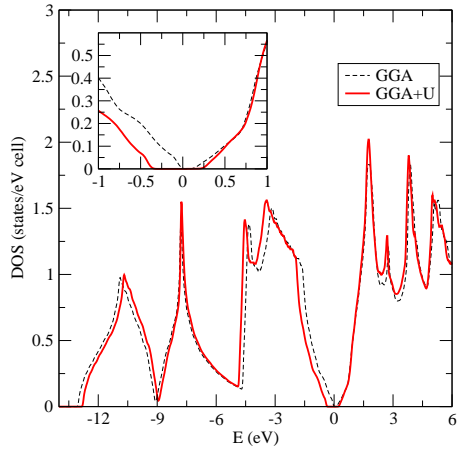}
\caption{Zincblende Ge density of states, with (red bold line) and without (black solid line) Hubbard U = 2eV on Ge $p$ states.} \label{fig:gedos}
\end{SCfigure}

In Fig. \ref{fig:gedos} we report the density of states of pure Ge
with and without the introduction of the Hubbard U. As expected, the occupied Ge $p$ band is pushed at lower energies and, as a side effect, the Ge $s$ and 3$d$ bands become slightly more localized, giving an overall smaller valence band width (by about 1-2 eV upon introducing U).

\section{Hybridization mechanism between Mn $d$ states and semiconductor valence band}
\label{azlev}
\begin{SCfigure}[4][b]
\includegraphics[width=.53\textwidth]{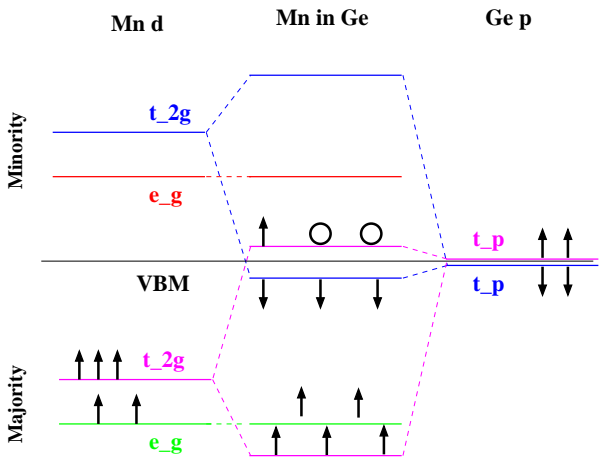}
\caption{Schematic diagram of Mn $d$ (right, splitted by exchange and crystal field effects) and Ge $p$ (left)
 atomic levels and how they hybridize when Mn is introduced in a Ge matrix. Arrows denote the direction of spins, whereas empty circles denote the holes.} \label{fig:lev}
\end{SCfigure}

In this section, we recall the hybridization mechanism between Mn atomic levels 
and SC states\cite{az:levels}. In Fig.\ref{fig:lev} we show the Mn $d$ atomic levels (left side), that are splitted by the exchange interaction and by a cubic tetrahedral crystal field.  By symmetry,  the resulting double-degenerate
$e_g$ levels  cannot efficiently hybridize with the host states and remain more or less unaltered in the DMS (see green and red states for majority and minority spins, respectively). On the other hand, the triply-degenerate $t_{2g}$ levels have both a symmetry and an energy-position that allow a strong hybridization with atomic Ge $p$ states
(shown on the right side). Therefore, a bonding-antibonding pair is formed in both majority and minority spin channels (see magenta and blue states, respectively). Also shown is the SC valence band maximum (VBM), which in the unperturbed host predominantly has an anion $p$ character. Now, when Mn (3$d^5$ 4$s^2$) substitutes a Ge (4$s^2$ 4$p^2$) atom, the total number of electrons that participate in the bonding includes
7 Mn  electrons (5 $d$ + 2 $s$) and 2 Ge $p$ electrons (Ge $s$ states have a much lower energy). The filling of the relevant levels by 9 electrons (starting of course from the bottom) shows that {\em i}) there are 6 and 3 electrons in the occupied valence band in the majority and minority spins, respectively , so that the total magnetic moment is 3 $\mu_B$; {\em ii}) the triply degenerate antibonding state in the majority channel is occupied by only one electron, therefore giving rise to two holes (see empty circles).
Therefore, Mn in Ge acts as a source of local magnetic moment and as a double acceptor.
A similar mechanism acts in GaMnAs, the only exception being that in a III-V SC there is one electron more coming from As (with respect to Ge); therefore, in the GaMnAs case, the $t_{2g}^{\uparrow}$ antibonding state 
is occupied by two electrons, giving rise to a total magnetic moment of 4 $\mu_B$ and to a single hole.

An additional point is worthwhile remarking: The $p-d$ hybridization pushes down (up)
the $p$ states of the anion with respect to the unperturbed host in the minority (majority) spin channel, resulting in an exchange-splitted VBM. In turn, there is a band-gap opening
(basically between the $t_{2g}^{\downarrow}$ bonding and the $e_g^{\downarrow}$ states) that gives rise to the celebrated half-metallicity ({\em i.e.} carriers at the Fermi level show a 100 $\%$ spin-polarization).  Now, it is well known that Ge  turns out to be a {\em semimetal} within DFT, {\em i.e.}  conduction states around the X point of the BZ have a lower energy than the VBM at $\Gamma$; this is of course due to acknowledged failures of DFT in describing excited states. However, despite the {\em metallic} character of the unperturbed Ge host,  due to the hybridization mechanism above, upon Mn-doping, a gap appears in the minority channel of MnGe, which shows a 100 $\%$ spin polarization at the Fermi level \cite{stroppa,zhao}.

\section{\bf Mn/Ge digital alloy:  band-gap effects}
\label{band-gap}
\subsection{Electronic structure}

\begin{SCfigure}
\includegraphics[width=.45\textwidth]{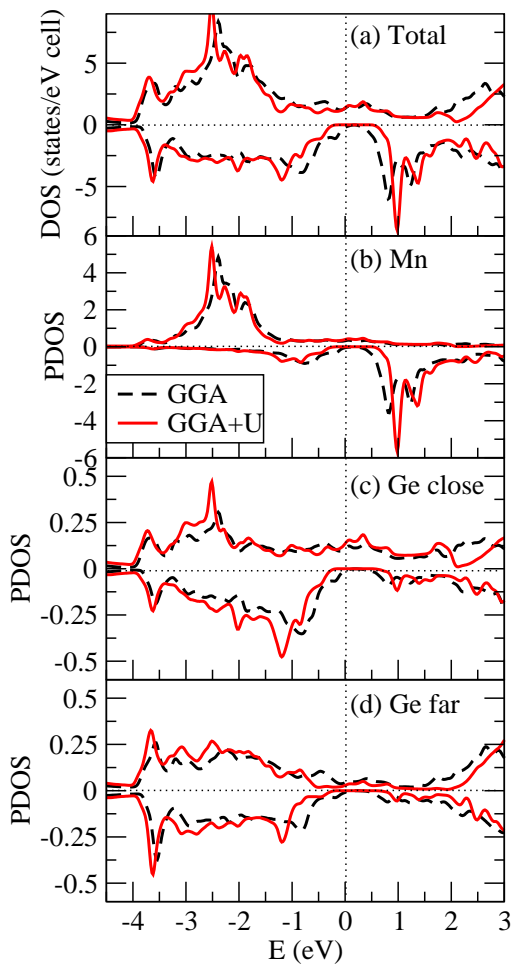}
\caption{Density of states for the Mn/Ge digital alloy. Panel (a): Total DOS. Panels (b), (c) and (d) show the PDOS of Mn, its nearest-neighbor Ge atom and a Ge far from Mn, respectively. The bold red  (dashed black) line shows the results with (without) the
introduction of U on Ge.} \label{fig:dos}
\end{SCfigure}

We now turn to the comparison between the electronic structure
obtained with and without the U on Ge.
In Fig. \ref{fig:dos} we show the total density of states and that projected (PDOS) on the relevant atoms:
Mn, its first Ge nearest-nighbor  and a Ge far from Mn (located at (0,0,$a$) with Mn sitting in the origin). It is interesting to note that, upon introduction of U, similarly to what
happened in pure Ge, the band-gap in the minority spins opens up to a value of $\sim$ 1 eV. On the other hand, there are no effects in the occupied majority spin channel, where the perturbation is screened out
up  to the $t_{2g}^{\uparrow}$ antibonding level. However, due to hybridization mechanism explained above that leads to an upward (downward) shift of the SC band-gap for the majority (minority) spins, this same effect of ``band-gap" opening is evident at 1-2 eV above $E_F$ (see for example the much larger forbidden energy range in the Ge far from Mn, Fig. \ref{fig:dos} (d)). 

From the point of view of magnetic moments, however, the situation does not change much: the half--metallicity is kept and, therefore, the total moment of 3 $\mu_B$ does not change. Moreover, no charge
rearrangement occurs
in the majority states (at least up to $E_F$) and the changes in the minority states simply results in lowering the energies of those states that were occupied anyway, even without Hubbard U. As a result, the Mn magnetic moment and its first-nearest Ge neighbour keep having - even upon introduction of U - a magnetic moment of 3.29 $\mu_B$ and -0.17
$\mu_B$, respectively, consistent with what previously reported in the literature.\cite{stroppa}

\subsection{Exchange interactions}

\begin{SCfigure}[4][b]
\includegraphics[width=.53\textwidth]{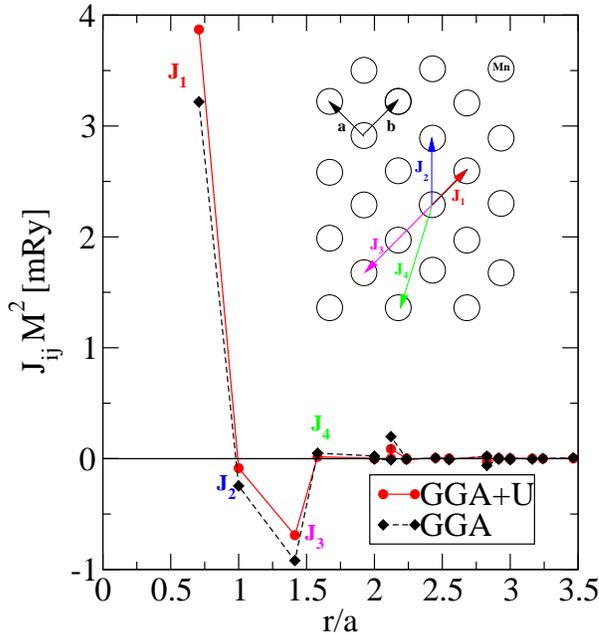}
\caption{Exchange constants multiplied by the squared magnetic moment (in mRy) for a Mn/Ge DA, as a function of the Mn-Mn distance (normalized to the lattice constant) .
Red circles (black diamonds) show the values for a larger (smaller) band-gap, achieved by means of GGA+U. In the inset, we show the in-plane fcc arrangement of Mn atoms (represented by empty circles): $J_1$ and $J_3$ are along the [110] direction, $J_2$ is along the [100] direction, whereas $J_4$ is along the [310] direction. Also shown in the insets are the in-plane {\bf a} and {\bf b} Bravais lattice vectors.} \label{fig:jconst}
\end{SCfigure}

The calculated exchange constants are shown in Fig.\ref{fig:jconst} for the Mn/Ge DA, with and without the introduction of the Hubbard parameter on the Ge $p$ states. In order to clarify
which are the relevant exchange interactions and along which crystalline direction they act, we also schematically show in the inset of Fig.\ref{fig:jconst} a top-view of the Mn plane in
the DA (with the Mn atoms arranged on a fcc lattice), along with the first few
calculated exchange interactions. 

Let us first discuss their general trends, irrespective of the effects of U.
Consistently with what previously found\cite{zhao,cinesi,stroppa}, we find that the first nearest-neighbor coupling $J_1$, directed
along the [110] axis where a strong $p-d$ bonding occurs, 
is strongly ferromagnetic (FM). On the other hand, 
 the second neighbor coupling $J_1$ along the [100] is small  and antiferromagnetic (AFM). Now, our calculated third-neighbor interaction, which again occurs along [110], is rather strongly AFM.
 This is at variance with what occurs in bulk MnGe, where the interaction between the two Mn atoms at the same distance
 of $a\sqrt(2)$ is FM, as previously reported\cite{zhao,cinesi}. However, we have to keep in mind that our geometry is rather
 different from that considered in Refs.\cite{zhao,cinesi}: there, the two Mn interact with no other Mn around or in-between, whereas here
there is another Mn in-between the two considered Mn atoms that are connected via $J_3$, along with other Mn atoms in the same plane that might also influence $J_3$. As a result, we get that the value of the exchange constants - at least for a high concentration of Mn - is rather sensibly affected by the magnetic environment.
As expected for half-metals, where the presence of a  minority band-gap causes a fast decay of exchange interactions as a function of distance, we find that interactions for Mn-Mn distances higher than $\sim$2$a$ are negligible. In addition, we remark that the interplanar interactions - which occurs at a large distance ({\em i.e.} 2$a$) and along the [100] direction for which also $J_2$ is already pretty small - are totally negligible, confirming the 2-dimensional character of the exchange-interactions in our DA model.

Let us now turn to the effects of a larger gap on the exchange constants. As a result of the introduction of U, the first neighbor interaction increases and also $J_2$ and $J_3$ become ``less AFM", therefore leading to an overall enhanced FM coupling. The presence of a higher gap is consistent with our findings of a larger ``damping" for exchange interactions, as a function of distance: 
interactions higher than $J_3$ are really negligible and, in any case, smaller in absolute value than those obtained within a bare-LSDA approach.

\section{Exchange constants: comparison between Mn/Ge and Mn/GaAs}
\label{chemical}
\begin{SCfigure}[4][b]
\includegraphics[width=.53\textwidth]{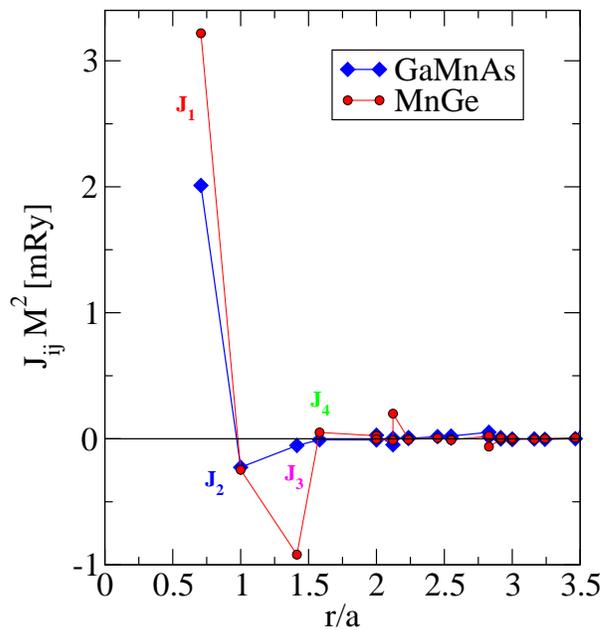}
\caption{Exchange constants multiplied by the squared magnetic moment (in mRy) as a function of the Mn-Mn distance (normalized to the lattice constant: comparison between Mn/Ge (red circles)
and Mn/GaAs (blue diamonds). Labels for exchange constants follow the notation shown in 
Fig. \protect\ref{fig:jconst}.} 
\label{fig:jconstgegaas}
\end{SCfigure}

In Fig.\ref{fig:jconstgegaas} we show our calculated exchange constants for Mn/Ge and Mn/GaAs, as a function of the Mn-Mn distance, in order to focus on
 the effects of the host on the exchange constants.
 As well known, Ge has a smaller gap and, obviously, a smaller ionicity than GaAs. This explains why the exchange constants fall off more rapidly in GaAs than in Ge as a function of the Mn-Mn distance.
Moreover, the first-neighbor $J_1$ is smaller in GaAs than in Ge, suggesting somehow a less efficient FM coupling. This is even strengthened by the real value of the exchange constant: in fact, we show in Fig.\ref{fig:jconstgegaas} the quantity $J_{ij}\cdot M^2$ and the values of the Mn magnetic moments
in Mn/Ge and Mn/GaAs are $\sim$3.2 and $\sim$3.7 Bohr magnetons, respectively, therefore $J_1$ in Mn/GaAs is almost half of its same value in Mn/Ge. 
However, when moving to $J_3$ and $J_4$ ($J_2$ has basically the same value in both compounds), the values for GaAs are ``less AFM" and FM, respectively, thus leading to a behaviour that compensates somehow the smaller FM $J_1$ coupling. Overall, the global behaviour should be similar in the two cases. Indeed, one could estimate the Curie temperature using a Montecarlo program along with a classical Heisenberg Hamiltonian; however, due to the two-dimensional nature of the systems,
which in principle prevents long-range FM according to the Mermin-Wagner theorem\cite{mermin}, one should include an anisotropy term - also calculated from first-principles -, which goes beyond the scope of the present work.

\section{Conclusions}

We presented first-principles calculations of the exchange constants and electronic structure
of Mn/Ge and Mn/GaAs magnetic semiconductor
``digital alloys". Upon increasing the band-gap in Mn/Ge digital alloys by means of the GGA+U approach, the half-metallicity is kept and the magnetic moments are almost totally unaffected.
However, the exchange constants for different Mn-Mn distances show an increased FM value for first-neighbors and have a weaker AFM character for higher-order neighbors, so as to induce an overall enhanced ferromagnetism when the minority band-gap is increased. Upon comparing Mn/Ge and Mn/GaAs, the exchange constant for first-neighbor coupling is surprisingly larger in Mn/Ge than in Mn/GaAs, whereas the couplings for larger Mn-Mn distances show a stronger ferromagnetic behavior in Mn/GaAs. Finally, in both cases where there is a larger gap (either obtained via GGA+U in Mn/Ge or obtained via chemical effects in Mn/GaAs), the exchange constants fall off much more rapidly, therefore giving rise to almost negligible exchange couplings for Mn-Mn distances larger than twice the lattice constant.

\end{document}